\documentstyle[pre,twocolumn,aps,epsf]{revtex}
\begin{document}
\draft
\twocolumn[\hsize\textwidth\columnwidth\hsize\csname@twocolumnfalse%
\endcsname

\title{Linear stability analysis of the Hele-Shaw cell
with lifting plates}

\author{S.-Z. Zhang \cite{now} and E. Louis}
\address{
Departamento de F{\'\i}sica Aplicada, Universidad de Alicante,\\
Apartado 99, E-03080 Alicante, Spain.}

\author{O. Pla and F. Guinea}
\address{
{I}nstituto de Ciencia de Materiales, Consejo Superior de Investigaciones
Cient{\'\i}ficas, \\ Cantoblanco, E-28049 Madrid, Spain.}

\date{\today}

\maketitle

\begin{abstract}
The first stages of finger formation in a Hele--Shaw cell with lifting plates
are investigated by means of linear stability analysis. The equation of motion
for the pressure field (growth law) results to be that of the directional 
solidification problem in some unsteady state. At the beginning of lifting
the square of the wavenumber  of the dominant mode results to be proportional
to the lifting rate (in qualitative agreement with the experimental data), to 
the square of the length of the cell occupied by the more viscous fluid,
and inversely proportional to the cube of the cell gap. This dependence on the
cell parameters is significantly different of that found in the standard cell.
\end{abstract}

]
\narrowtext
\section{Introduction}

Despite of the great effort devoted lately to improve the understanding of
Saffman-Taylor (ST) instabilities\cite{ST,SEVERAL}, many related experimental
facts still lack a sound explanation \cite{CLOUD}. Here we are interested in
the experimental observations on Hele-Shaw (HS) cells with lifting
plates\cite{JACOB}. This variation with respect to the standard constant
gap HS cell was suggested as a way to bring the ST problem closer to
directional solidification \cite{LANGER,KURZ}. In this experiment (see 
Fig.~\ref{f:helle-shaw}), instead of applying pressure to the less viscous
fluid, the upper plate is lifted at the less viscous side (commonly air)
at a fixed rate. It seems clear that the lifting of the upper plate will
promote a pressure gradient analogous to the temperature gradient present in
directional solidification. An interesting variation of this experiment is
the Hele--Shaw cell with a small gap gradient investigated by Zhao et
al\cite{ZHAO}.

The main results  of the work reported by Ben--Jacob et al\cite{JACOB}
concern the spacing of the dendrites. These authors observed that this spacing
decreased as the lifting velocity was increased. They also noted a dependence
on the initial plate spacing. On the other hand, although the experimental
results for the HS with lifting plates have been analysed by means of a
simplified version of the growth law \cite{ROCHE,LOUIS}, a detailed analysis
of this system is still lacking. 

In this work we present a study of growth instabilities in Hele--Shaw cells
with lifting plates. We first derive the basic equation (growth law), which
results to be different from that proposed by Ben--Jacob et al \cite{JACOB}.
In particular our equation is not  homogeneous, and thus the comparison with
the simpler version of the directional solidification problem is not so
evident. Then we discuss the boundary conditions and carry out the linear
stability analysis. Our results qualitatively explain the experimental
data reported by Ben--Jacob et al \cite{JACOB}.

\section{Basic equations and boundary conditions}

Flow in the Hele-Shaw cell is governed by the Navier-Stokes equation
\cite{LANDAU,BENSIMON}
  \begin{equation}
    \frac{\partial {\bf v}}{\partial t} +({\bf v}.{\bf \nabla}){\bf v}
    =-\frac{1}{\rho}{\bf \nabla}P+\frac{\mu}{\rho}\nabla^2{\bf v}\;,
  \end{equation}
where ${\bf v}$ is the speed of the fluid and $P$ the pressure. $\rho$ and
$\mu$ are the density and the viscosity of the fluid, respectively. For small
Reynolds numbers and assuming that the time derivative of the fluid velocity
is much smaller than its spatial derivatives, this equation reduces to
  \begin{equation}
    {\bf \nabla}P=\mu\nabla^2{\bf v}\;.
  \end{equation}

\begin{figure}
\epsfxsize 6cm
\epsfbox{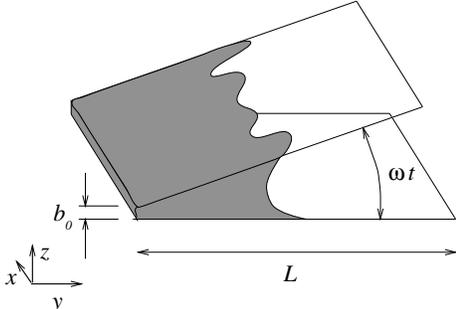}
\caption{Schismatic representation of the Helle-Shaw cell with lifting plates,
showing the parameters used in the text.
\label{f:helle-shaw}}
\end{figure}

Averaging over the the cell gap leads to the Poiseuille-Darcy equation
for the mean velocity of the fluid,
  \begin{equation}
    {\bf v}_j=-\frac {b^2}{12\mu_j}{\bf \nabla} P_j\;, 
  \end{equation}
where ${\bf v}_j$,  $\mu_j$, and $P_j$ are the velocity, viscosity 
and pressure field of fluid $j$ ($j$=1,2); and $b$ is the gap of the cell.
In order to obtain an equation for the pressure field we need to combine Eq.
(3) with mass conservation. In the present case the latter deserves a careful
consideration. Let the HS cell lie in the $x-y$ plane, the $y$--axis
being the direction of motion of the fluids, and the origin of coordinates
be at the closed (fixed) end of the cell (see Fig.~\ref{f:helle-shaw}). The
gap of the cell varies as
  \begin{equation} 
    b(y,t) = b_0 + y\,\tan(\omega t)\;, 
  \end{equation}
where $\omega$ is the lifting angular speed (we shall hereafter 
call $a=\,\tan(\omega t)$). As a consequence, the mass within a thin column 
of height $b$ changes as, $\delta m/\delta t \propto  \delta b /\delta t $
(where the density of the fluid $\rho$ is assumed to be constant). Then, the
equation which describes mass conservation reads
  \begin{equation}
    \frac{\partial b}{\partial t} = -{\bf \nabla}\cdot(b {\bf v}_j)\;. 
  \end{equation}

It should be here noted that mass (and density) conservation requires that,
neither bubbles are formed nor drops of the displaced fluid are left behind
during the lifting process. This condition, although unlikely accomplished in
actual experiments, it is probably unavoidable in analytical calculations.
Eq. (5), combined with the Poiseuille-Darcy equation (Eq. (3)), gives the
differential equation (growth law) which governs flow in the HS cell with
lifting plates,
  \begin{equation}
    \nabla^2 P_j+\frac {3a}{b} \frac { \partial P_j}{ \partial y }= 
    \frac {12\mu_j \dot{a} y}{b^3}\;,
  \end{equation}
where $\dot{a}$ expresses the derivative of $a$ with respect to time. It 
is interesting to note that this is an inhomogeneous equation as opposed
to the homogeneous one reported in Ref. [4]. The inhomogeneous term comes
from the time dependence of the gap (Eq. (4)) and it is not present in the
growth law for the HS with a gap gradient, as already found by Zhao et 
al\cite{CLOUD,ZHAO}. Note, however, that the inhomogeneous term
is essential in this case, as no fluid is injected at the open end of the cell
to promote fluid motion (see below). We also note that including the spatial
dependence of the gap $b$ in Eq. (5) leads lo a factor 3 in the second term of
Eq. (6), as opposed to the factor 2 reported in Ref. [4] and in agreement with
Zhao et al\cite{ZHAO}. This change in the growth law makes
calculations slightly more complicated. As regards the comparison with the
directional solidification problem, we note that Eq. (6) is similar to that
which describes directional solidification in some
{\sl unsteady}-state\cite{KURZ}. 

In discussing the boundary conditions let 
$L$ be the length of the cell and $L_i$ the length of the zone occupied
by the more viscous fluid at $t$=0.
On the other hand, and in order to carry out the linear stability analysis, 
we assume that the interface between the two fluids, instead of being flat,
is slightly perturbed as $y_p=y_i+\delta {\rm e}^{iqx}$, where 
$\delta=\delta_0{\rm e}^{\gamma t}$ is the time--dependent amplitude of the 
perturbation (assumed to be much smaller than $y_i$) and $q$ its wavenumber. 

The  first boundary condition accounts for the fact that the 
fixed end of the cell ($y=0$) is closed, 
  \begin{equation}
    v_{1y}(0)=0\;.
  \end{equation}
On the other hand, the velocities of the two fluids must be equal at the
interface 
  \begin{equation}
    v_{1y}(y_{p})=v_{2y}(y_{p})\;.
  \end{equation}
Finally the forces which the fluids exert on each other at the common
interface must be equal and opposite. This condition can be
approximated by
  \begin{equation}
    P_1(y_p)-P_2(y_p)  = 
    \sigma \left[\frac {1}{R_x(y_p)} + \frac {1}{R_z(y_p)} \right]\;, 
  \end{equation}
where $\sigma$ is the surface-tension (the interfacial tension associated
to the interface between the two fluids) and $ R_x$, $R_z$, the principle
radii of curvature of the interface at a given point,
\begin{mathletters}
  \begin{equation}
    \frac {1}{R_x(y_p)} =- \frac {\partial^2 y_p}{\partial x^2 }\;,
  \end{equation}
  \begin{equation}
    \frac {1}{R_z(y_p)} \approx  \frac {1}{b(y_i,t)}\;.
  \end{equation}
\end{mathletters}

The latter equation is related to wetting effects. Although, as already
reported by other authors\cite{ZHAO,BENSIMON,PARK}, we have found that these
effects give a negligible contribution, we have kept this term all throughout
the calculation.

\section{linear stability analysis}

Once the plane interface is perturbed, the most general solution of Eq.(6)
can be written as 
  \begin{equation}
     P_j(x,y)=f_j(y)+g_j(y) e^{iqx}\;,
  \end{equation}
where $g_j(y)$ is proportional to the amplitude of the perturbation $\delta$.
Introducing Eq. (11) into the growth law (Eq. (6)) we obtain, 
\begin{mathletters}
  \begin{equation}
     \frac{\partial^2  f_j}{\partial y^2}+\frac{3a}{b}\frac{\partial f_j}
     {\partial y} = \frac{12\mu_j\dot{a}y}{b^3}\;,
  \end{equation}
  \begin{equation}
     \frac{\partial^2 g_j}{\partial y^2}+\frac{3a}{b}\frac{\partial g_j}
     {\partial y} - q^2 g_j= 0\;. 
  \end{equation}
\end{mathletters}

The boundary conditions for the fluid velocity at the closed end give
\begin{mathletters}
  \begin{equation}
    \left.\frac{\partial f_1}{\partial y}\right|_{y=0}=0\;,
  \end{equation}
  \begin{equation}
    \left.\frac{\partial g_1}{\partial y}\right|_{y=0}= 0\;.
  \end{equation}
\end{mathletters}
Note that in order to obtain physically meaningful solutions
we should also require that the perturbation be damped in the
open end of the cell, this means that
$(\partial g_2/\partial y)\vert_{y=L}=0$. This assumption
will be valid whenever the viscosity $\mu_2$ is very small (note that,
in the present work, we will analyze experiments in which fluid 2 is air). 
The continuity of the velocity at the interface between the two fluids 
leads to (up to first order in $\delta$):
\begin{mathletters}
  \begin{equation}
    \frac{1}{\mu_1}\left.\frac{\partial f_1}{\partial y}\right|_{y=y_i}=
    \frac{1}{\mu_2}\left.\frac{\partial f_2}{\partial y}\right|_{y=y_i}\;,
  \end{equation}
  \begin{equation}
    \frac{1}{\mu_1}\left(\left.\delta\frac{\partial^2 f_1}{\partial y^2}
    +\frac{\partial g_1}{\partial y}
    \right)\right|_{y=y_i}\!\!\!\!\!\!=\frac {1}{\mu_2}\left.\left(\delta 
    \frac{\partial^2 f_2}{\partial y^2}+\frac{\partial g_2}
    {\partial y}\right)\right|_{y=y_i}\!\!\!\!\!\!\!.
  \end{equation}
\end{mathletters}
Finally, the continuity of forces at the interface gives
\begin{mathletters}
  \begin{equation}
     \left [f_1-f_2-\frac{\sigma}{R_z} \right ]_{y=y_i} =0\;,
  \end{equation}
  \begin{equation}
    \left[ g_1-g_2(y) \right]_{y=y_i}\!\!\!=-\left[\frac{\partial\left(
    f_1-f_2\right)}{\partial y}\right]_{y=y_i}\!\!\!-\sigma q^2\;. 
  \end{equation}
\end{mathletters}

The solution of Eq. (12a), using the boundary conditions Eqs. (13a) and
(14a) can be written as, 
  \begin{equation}
    f_j=\frac{12\mu_j\dot{a}}{a^3}\!\left(\frac{b_0}{b(y,t)}\!+\!\frac{1}{2}
    \ln b(y,t)\!-\!\frac {b_0^2 }{4b(y,t)^2}\right)\!+\!A_j, 
  \end{equation}
where $A_j$ ($j$=1,2) are constants that can be determined from the 
boundary condition of Eq. (15a) (we will not need them here). It should be
noted that, had Eq. (6) been homogeneous, the solution of Eq. (12a) would
have been, $f_j=A_j-B_j/b(y,t)^2$, which, after proper use of the boundary
conditions, gives a uniform pressure field and, therefore, no fluid motion,
as remarked above. On the other hand the solution of Eq. (12b) is
  \begin{equation}
    g_j=\frac {1}{\xi} \left [C_jI_1(\xi)+D_jK_1(\xi) \right ]\;, 
    \hspace{1.cm}  j=1,2\;,
  \end{equation}
and its derivative,
  \begin{equation}
     \frac {\partial g_j}{\partial y} = \frac {q}{\xi} \left 
     [C_jI_2(\xi)-D_jK_2(\xi) \right ]\;, \hspace{1.cm}  j=1,2\;,
  \end{equation}
where $ \xi=q b(y,t)/a $, and $ I_1(\xi) $, $I_2(\xi)$, and $K_1(\xi) $,
$K_2(\xi)$ are the first and the second order modified Bessel functions.
The constants $C_j$, $D_j$ have to be determined from the boundary conditions, 
\begin{mathletters}
  \begin{equation}
     C_1I_2(\xi_0)-D_1K_2(\xi_0)=0\;,
  \end{equation}
  \begin{equation}
     C_2I_2(\xi_L)-D_2K_2(\xi_L)=0\;,
  \end{equation}
  \begin{equation}
    \left(\frac {C_1}{\mu_1}-\frac {C_2}{\mu_2}\right)I_2({\xi_i})-
    \left(\frac {D_1}{\mu_1}-\frac {D_2}{\mu_2}\right)K_2({\xi_i})=0\;,
  \end{equation}
  \begin{equation}
    (C_1-C_2)I_1(\xi_i) +(D_1-D_2)K_1(\xi_i)= \epsilon \delta\;,
  \end{equation}
\end{mathletters}
where $\xi_0=qb_0/a $, $\xi_i= q(b_0+ay_i)/a $, $\xi_L =q(b_0+aL)/a$, and
$\epsilon$ is given by,
  \begin{equation}
    \epsilon =\xi_i \left[\frac {12(\mu_1-\mu_2)V_i(y_i)}{b(y_i,t)^2}+
    \sigma q^2 \right ]\;,
  \end{equation}
where the velocity of the interface between the two fluids is given by,
  \begin{equation}
    V_i(t) = -\frac{b^2}{12\mu_j}\left.\frac{\partial f_j}{\partial y}
    \right|_{y=y_i}\!\!\!={\dot y}_i=-\frac{{\dot a}y_i^2}{2b(y_i,t)}\;,
  \end{equation}
independent of $j$. Note that, due to the choice of the origin of coordinates,
this velocity is negative. Integrating this velocity gives an expression for
the position of the interface (where the boundary $y(t=0)=L_i$ is taken):
  \begin{equation}
    y_i(t) = \frac{\sqrt{b_0^2 + 2L_i b_0 a} - b_0}{a}
    \;.
  \end{equation}
which is the same result as if we had considered the simpler mass conservation
as $\frac{1}{2} ay_i^2 + b_0 y_i = b_0 L$.

Now, we have all the ingredients to calculate the instantaneous velocity of
the interface (growth rate), which can be calculated from the Poiseuille-Darcy
equation. The result is,
  \begin{eqnarray}
    v_{1y}(y_p)&=&\dot{y_p} = \dot{y_i}+\dot{\delta}{\rm e}^{iqx} \nonumber \\
    &=&-\frac {b}{12 \mu_1}\left.\frac{\partial(f_1+g_1{\rm e}^{iqx})}
    {\partial y}\right|_{y=y_p}\;.  
  \end{eqnarray}
From which we obtain the velocity of the perturbation $\gamma$, 
  \begin{equation}
     \gamma=\frac {\dot{\delta}}{\delta}=-\frac { \dot{b}(y_i,t)}{b(y_i,t)} 
     - \frac {a b(y_i,t) \epsilon}{12}Z(\xi_i,\xi_L)\;, 
   \end{equation}
where,
  \begin{eqnarray}
    Z(\xi_i,\xi_r)=\frac {I_2(\xi_i)}{I_1(\xi_i)}&\times &
    \left(\mu_1 \frac{\left[ G_2(\xi_0)+G_1(\xi_i)\right]}
    {\left[ G_2(\xi_0)-G_2(\xi_i) \right]}\right.
    \nonumber \\&&   
    \left.-\mu_2 \frac{\left[ G_2(\xi_L)+G_1(\xi_i) \right ]}
    {\left[ G_2(\xi_L)-G_2(\xi_i) \right]} \right)^{-1}\!\!\!, 
  \end{eqnarray}
and,
  \begin{equation}
     G_j(\xi)=K_j(\xi)/I_j(\xi)\;, \hspace{1cm} j=1,2\;.
\end{equation}

At the very beginning of liftting $1 << \xi_0 \alt \xi_i \alt \xi_L$.
As a consequence, $ G_1(\xi_i) \simeq G_2(\xi_i) $. On the other hand, and
due to the exponential behavior of the Bessel and Hankel
functions, $ G_2(\xi_0) >> G_2(\xi_i) >> G_2(\xi_L)$. Thus, Eq. (23b)
can be approximated as 
  \begin{eqnarray}
     \gamma&=&-\frac { \dot{b}(y_i,t)}{b(y_i,t)} \nonumber\\&& 
     -\frac {b(y_i,t)^2}{\mu_1+\mu_2}q \left [ \frac {12(\mu_1-\mu_2)
     V_i(t)}{b(y_i,t)^2}
     +\sigma q^2 \right ] \;.
  \end{eqnarray}

The wavenumber of the dominant mode $q_d$ results to be,
  \begin{equation}
     q_d^2= \frac {4(\mu_1-\mu_2)V_i(t)}{\sigma b(y_i,t)^2}\;. 
  \end{equation}
This result is quite similar to that obtained for the standard Hele-Shaw 
cell. The only difference resides upon the fact that the interface
velocity $V_i(t)$ and the gap $b(y_i,t)$ now depend on time.  
As regards the cutoff wavenumber, we note that if the first
term in the r.h.s of Eq. (27) is neglected (in fact it is quite small
for most of the experimental configurations and conditions), the result
is again equivalent to that of the standard cell ($q_c^2=3q_d^2$). If it
is not neglected, a minimum wavenumber, below which the
system is stable, is also found. 

At $t=0$, the dominant mode is
  \begin{equation}
     q_d^2= \frac {4(\mu_1-\mu_2)}{\sigma b_0^2} V_i(0)\;,
  \end{equation}
where $V_i(0)$ is the velocity of the interface at $t=0$,
  \begin{equation}
     V_i(0)= \frac{ \omega L_i^2}{2b_0}\;.
  \end{equation}
$V_i(0)$ introduces a dependence on the cell parameters 
of the wavenumber of the dominant mode, not present
in the standard cell. In particular, $q_d$ depends on the square of
the length occupied by the displaced (more viscous) fluid. It
is also inversely proportional to the gap, leading to a $b_0^{-3}$
dependence of $q_d$, as opposed to $b_0^{-2}$ in the standard cell. 
These differences could be easily checked experimentally.

\section{Numerical Results and Discussion}

At the beginning of lifting ($t=0$) the process is governed by Eq. (30).
In actual experiments, as the less viscous fluid is commonly air,
$\mu_2\approx0$,
and the wavelength of the dominant and the cutoff modes are given by
expressions identical to those for the standard cell, namely,
$\lambda_d/b_0 = \pi(\sigma/\mu_1V_i(0))^{1/2}$, and
$\lambda_c = \lambda_d/\sqrt{3}$, respectively. 
In order to estimate the wavelength of the dominant mode we
consider the experimental set up investigated by Zhao et
al \cite{ZHAO}. These authors used air/glycerine ($\mu$ = 
65 mPa and $\sigma$=29.5 N/m) in a cell with a gap of 2.5 mm and 
a length of 1 m (we assume that glycerine fills the whole cell).
Taking $\omega$ = 0.001 rad/s, $\lambda_d\approx$ 1.2 cm. This
result decreases in an order of magnitude if, as done by
Ben--Jacob et al \cite{JACOB}, the lifting rate is increased
in two orders of magnitude. We cannot precisely compare our results
with the data obtained by the latter authors as they do not give important
parameters such as the length and gap of the cell.

\begin{figure}
\begin{picture}(250,180) (-10,-15)
\epsfbox{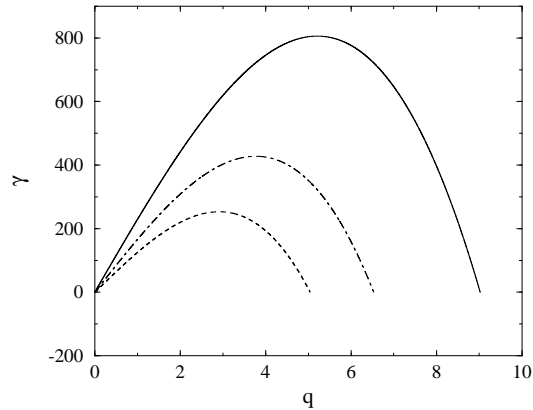}
\end{picture}
\caption{
Results for the velocity of the perturbation $\gamma$ (rad/s) as a function
of its wavenumber $q$ (cm$^{-1}$), for several times after lifting 
is initiated:
$t$=0 s (continuous line), $t$=0.5 s (chain line) and $t$=1 s (dashed line)
. The results correspond to the cell parameters
of Ref. [7], namely:  air/glycerine ($\mu$ = 
65 mPa and $\sigma$=29.5 N/m) in a cell with a gap of 2.5 mm and 
a length of 1 m (we assume that glycerine fills the whole cell).
The lifting rate is $\omega$ = 0.001 rad/s. 
\label{fig.1}}
\end{figure}

If instead of glycerine we consider water, at $t=0$ and for
the same cell parameters, the wavelength of the dominant mode results
to be $\approx$ 0.4 cm$^{-1}$, which leads to a wavelength of 
$\approx$ 16 cm. This wavelength is of the order of the full
width of standard cells and therefore one should not expect
the formation of fingers. Further we note that the velocity
of the perturbation results to be at least an order of
magnitude smaller than for glycerine, reinforcing our view that
the flat front would be rather stable. Note that
this result is also obtained for the standard cell and that it
is in agreement with the experimental observations. The only way to increase 
the tendency towards instabilities would be to increase the
velocity of the interface which in the present case
can be accomplished by increasing $L$ and
the lifting rate and/or decreasing $b_0$. 

As lifting proceeds, fingers are being formed and 
a linear stability analysis such as that carried out here is 
no longer strictly
valid. However, some information valid to understand specific
features of the growing process may be obtained from
the results of a linear stability analysis, as already
found in the case of growth in systems governed by Poisson's 
equation \cite{LOUIS}.
In Figure 2 we report our results for the velocity at which the
perturbation propagates  as a function of the perturbation
wavenumber, for several times after lifting was initiated. In the
calculations the first term in the r.h.s. of Eq. (27) was
neglected. The results correspond to
the cell parameters given in the preceeding paragraph and
a lifting rate of $\omega$=0.001 rad/s. It is noted that the
velocity of the perturbation $\gamma$ strongly decreases with time. In
fact the value at its maximum is reduced in more than a factor of
5 after lifting the cell for 1 second. The wavanumber at which $\gamma$ shows
a maximum (dominant mode) and cutoff wavenumber do also
decrease with time (see Fig. 3). The decrease
of the velocity of propagation of the perturbation suggest a
lower tendency towards instabilities and, thus, a larger fractal
dimension of the aggregates (this is compatible with a decreasing $q_d$).
A variation in the fractal dimension
of the aggregates as growth (lifting) proceeds was also found by Roche et
al \cite{ROCHE} in their simulations of the Poisson equation,
which they argue to be adequate for the HS with lifting plates (note
that this is approximately correct if the term proportional to
the first derivative of the pressure in Eq. (6) is either
neglected or replaced by a constant).
However they found that the simulation which seemed to describe
more closely the experimental situation gave an increasing fractal 
dimension.  

\begin{figure}
\begin{picture}(250,180) (-10,-10)
\epsfbox{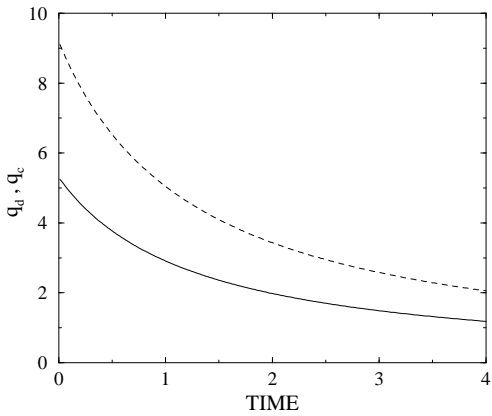}
\end{picture}
\caption{
Wavenumbers (cm$^{-1}$) of the dominant ($q_d$, continuous line) and cutoff 
($q_c$, broken line) modes as a
function of time (s) in the Hele--Shaw with lifting plates investigated
in this work. The results correspond to the parameters given
in the caption of Fig. 1.
\label{fig.2}}
\end{figure}

Summarizing, we have presented a linear stability study of the HS with
lifting plates. We have first derived the basic equations which result
to be that of the directional solidification problem under some
unsteady conditions. Despite the simplifications made to do the analytical
work (we do not treat completely the wetting effects and neglect the fluid that
stays attached to the plates), the results for the wavelength of the dominant
mode 
seem to be compatible with the available experimental data. It
will be worth carrying out more experiments in order to get
more information on the dependence of the characteristics of
the growing aggregates on the parameters of the cell. 

\acknowledgments
We wish to acknowledge financial support from  the spanish CICYT
(Grants No. MAT94-0058 and MAT94-0982). S.-Z. Zhang wishes to thank 
the "Ministerio
de Educaci{\'o}n y Ciencia" (Spain) for partial support.

\end{document}